\documentstyle[amsfonts,12pt]{article}

\def\half{\textstyle{\frac{1}{2}}}

\def\ra{\rightarrow}

\def\n{\nonumber}
\def\b{\begin{eqnarray*}}                  
\def\e{\end{eqnarray*}}                    
\def\bn{\begin{eqnarray}}                  
\def\en{\end{eqnarray}}                    
\def\<{\langle}
\def\tint{{\textstyle\int}}
\def\p{\phi}
\def\>{\rangle}

\def\v{\varphi}

\def\{{\lbrace}
\def\}{\rbrace}
\def\L{\Lambda}

\title{Nonrenormalizability and Nontriviality}
\author{John R. Klauder\\
Departments of Physics and Mathematics\\
University of Florida\\
Gainesville, Fl  32611}
\date{}                               
\begin{document}
\maketitle
\begin{abstract}
A redesigned starting point for covariant $\phi^4_n$, $n\ge 4$,  models 
is suggested that takes the form of an alternative lattice action and which 
may have the virtue of leading to a nontrivial quantum field theory in the 
continuum limit. The lack of conventional scattering for such theories is 
understood through an interchange of limits.
\end{abstract}
Despite being perturbatively nonrenormalizable, the quantum theory of 
covariant scalar $\phi^4_n$ models has been shown to be trivial for all 
space-time dimensions $n\ge5$, while for $n=4$ it is widely believed to 
be trivial as well \cite{fro}. Triviality follows by showing that the 
conventionally lattice-regularized, Euclidean-space functional integral 
tends to a Gaussian distribution in the continuum limit independent of 
any choice of renormalizations for the mass, coupling constant, and field 
strength. Although mathematically sound, a trivial result is inconsistent 
in the sense that the classical limit of the quantized theory differs 
from the original (nontrivial) classical theory. In this Letter we 
reexamine this problem once again, and suggest an alternative formulation 
whereby quantum models for $\phi^4_n$ may be nontrivial. Generally, in 
what follows, we set $\hbar=1$.

We start with a lattice-regularized, Euclidean-space functional integral 
expressed as
  \bn &&\hskip-1.1cm S_a(h)\equiv\<\exp(\Sigma\, h_k\phi_k a^n)\> \n \\
    &&\equiv N_a\int\exp\{\Sigma\, h_k\p_ka^n-\half Z\,\Sigma(\p_{k^*}
- -\p_k)^2a^{n-2}-\half Zm_o^2\,\Sigma\,\p^2_ka^n \n\\
  &&\hskip2.5cm-Z^2g_o\,\Sigma\, \p^4_ka^n-\Sigma\,P(Z^{1/2}\p_k,a)\,a^n\}
\,\Pi\,d\p_k\;. \en
Here $k=(k^0,\cdots ,k^{n-1})$ , $k^j\in {\mathbb Z}$ for all $j$, labels a 
lattice site; $k^*$ signifies one of the $n$ nearest neighbors to $k$ and 
the sums run over a large but finite hypercubic lattice; $a$ represents 
the lattice spacing; $h_k$ denotes the lattice-cell average of a smooth 
source function $h(x)$, $x\in {\mathbb R}^n$; $Z>0$, $m_o^2$, and 
$g_o\ge 0$ are functions of the cutoff $a$; and---for the present---the 
auxiliary term $P\equiv0$. We choose $N_a$ such that $S_a(0)=1$, and 
let $\<(\cdot)\>$ denote an average with respect to the resultant 
probability distribution. The continuum limit is defined as the limit 
$a\ra0$ in conjunction with a diverging number of lattice sites so that 
the space-time volume itself eventually tends to ${\mathbb R}^n$ in a 
suitable way. For $n\le3$, the continuum limit leads to acceptable 
(nontrivial) results \cite{gli}; for $n\ge5$, on the other hand, the 
continuum limit has the form \cite{fro}
  \bn \lim S_a(h)=\exp[\,\half\tint h(x)C(x-y)h(y)\,d^n\!x\,d^n\!y\,] \en
for a suitable covariance function $C(x-y)$, and all indications point to 
the same conclusion when $n=4$. [If $C(x-y)$ is not locally integrable, then 
this condition is replaced by one in terms of correlation functions with
noncoincident points.]

Let us sketch one plausible argument that leads to trivial behavior. 
Consider the dimensionless and rescaling invariant correlation-function 
ratios, which also admit meaningful continuum limits, given, for $r\ge 1$, by
\bn g_{(r)}\equiv\frac{\Sigma\<\p_0\p_{k_2}\cdots\p_{k_{2r}}\>^T}
{[\Sigma\<\p_0\p_k\>]^r[\Sigma k^2\<\p_0\p_k\>/
6\Sigma\<\p_0\p_k\>]^{n(r-1)/2}}\;;  \en
by symmetry, all odd-order correlations vanish. For $n\ge5$, 
mean field theory is generally accepted, and for small $a$ it 
leads to the behavior that
$g_{(r)}\propto a^{(n-4)(r-1)}$. Thus for $n\ge5$ and $r\ge2$, 
$g_{(r)}\ra0$ in the continuum limit. The Lebowitz inequality 
\cite{gli}, which states that $\<\p_j\p_k\p_l\p_m\>^T\le0$ for 
such models, linked with $g_{(2)}\ra0$, implies the vanishing of 
the truncated $(T)$ four-point function in the continuum limit. 
Among the class of distributions considered, only a Gaussian 
distribution admits a vanishing truncated four-point function, and 
triviality follows.
For $n=4$, logarithmic corrections to mean field theory arise and for 
small $a$ it is plausible that $g_{(r)}\propto|\ln(a)|^{-(r-1)}$, 
which again leads to triviality as $a\ra0$.

For $n\ge5$, the indicated $a$ dependence of $g_{(r)}$ stems from the 
separate mean field behavior, valid for small $a$, given by 
$\Sigma\<\p_0\cdots\p_{k_{2r}}\>^T\propto a^{-2-6(r-1)}$ and 
$\Sigma k^2\<\p_0\p_k\>\propto a^{-4}$ \cite{fis}. This behavior 
reflects the divergence arising from the (multiple) {\it sum} and is 
based, for a suitable choice of $Z$, on a largely $a$-{\it in}dependent 
correlation function $\<\p_0\cdots\p_{k_{2r}}\>^T$.  

Let us return to (1), now allowing for an auxiliary contribution 
$P\not\equiv0$. Normally, $P$ is determined by a perturbation analysis 
so as to cancel unwanted ultraviolet divergences in the continuum limit. 
Instead, we propose to {\it design} $P$ to achieve nontriviality. In 
particular, we choose $P$ so that the truncated correlation functions 
themselves become {\it uniformly} $a$-dependent, specifically, for all 
$r\ge1$ and $n\ge5$, that
\bn  \<\p_0\p_2\cdots\p_{k_{2r}}\>^T\propto a^{n-4}\;;  \en
for $n=4$ the right side of (4) should be replaced by $|\ln(a)|^{-1}$. 
With such a choice for $P$, the combined effects of the divergence due 
to long-range order and rescaled amplitude lead to $g_{(r)}\propto a^0=1$ 
for all $r\ge 1$, and for any $n\ge 4$. Choosing $Z\propto a^{n-4}$ 
[or $|\ln(a)|^{-1}$] properly rescales the correlation functions to 
macroscopic values. Such a theory would be non-Gaussian, hence 
nontrivial, in the continuum limit.

Based on experience with related but soluble models \cite{klb}, we 
conjecture that a $P$ of the form
\bn  P(\p_k,a)=A(a)\,\frac{[\p_k^2-B(a)]}{\,[\p_k^2+C(a)]^2}\;,  \en
for suitably chosen $A$, $B$, and $C$ (which also depend on $n$), may do 
the job. As $a\ra0$, we expect that $A\ra\infty$, $B\ra0$, and $C\ra0$. 
Thus the indicated expression is a regularized form of a formal continuum 
potential proportional to $1/\p(x)^2$. However, just as the $1/r^2$ 
potential that arises from the kinetic energy in a spherically symmetric, 
quantum-mechanical situation necessarily carries a proportionality factor 
of $\hbar^2$, it is more proper to recognize that the formal auxiliary 
potential $P$ is proportional to $\hbar^2/\p(x)^2$, i.e., $A\propto \hbar^2$. 
Carrying the analogy further, we observe that $P$ is not a counterterm 
for the quartic interaction but rather for the kinetic energy term. Thus 
we are led to propose that $P$ is a {\it nonclassical, auxiliary potential}, 
which explains its absence in a strictly classical limit in which 
$\hbar\ra0$. Based on related models, we are also led to conjecture 
that a $P$ having the desired properties leads to a quantum theory the 
classical limit of which agrees with the classical theory with which one 
started.

Assuming that some such $P$ exists, we can proceed to derive, in a general 
fashion, certain additional facts. The nature of the nonclassical, 
auxiliary potential $P$ leads to a generalized Poisson distribution in the 
continuum limit. In Minkowski space-time the operator structure of such 
fields admits a relatively simple superstructure. Introduce the basic 
(Fock) operators
$A_l$ and $A^\dagger_l$, $l\in\{0,1,2,\ldots\}$, where $[A_l,A_m]=0$ and 
$[A_l,A^\dagger_m]=\delta_{lm}$ for all $l$ and $m$, and $A_l|0\>=0$ for 
all $l$, with $|0\>$ unique. As usual, the Hilbert space is spanned by 
vectors of the form $|0\>,\;A^\dagger_l|0\>,\;A^\dagger_lA^\dagger_m|0\>$, 
etc. Furthermore, let $\L_{lm}(x)\;[=\L_{ml}(x)^*]$ and $C_l(x)$ denote a 
suitable set of complex fields. With summation implied, the Minkowski 
field operator has the representation \cite{kl2}
\bn  \varphi(x) =A^\dagger_l\,\L_{lm}(x)\,A_m+A^\dagger_l\,C_l(x)+C^*_l(x)
\,A_l\;. \en
It follows from (6) that
\bn &&\hskip-1cm\<0|\varphi(x_1)\varphi(x_2)\cdots\varphi(x_{2r-1})
\varphi(x_{2r})|0\>^T\n\\
&&\hskip1cm =C^*_{l_1}(x_1)\L_{l_1\,l_2}(x_2)\cdots \L_{l_{2r-2}
\,l_{2r-1}}(x_{2r-1})C_{l_{2r-1}}(x_{2r})\;, \en
assuming all odd-order correlation functions vanish. 
Since $\<0|\varphi(x)\varphi(y)|0\>=C^*_l(x)\,C_l(y)$, it follows, 
for example, from the spectral representation for $n\ge4$, that 
$\Sigma|C_l(x)|^2=\infty$.
The field $\varphi(x)$ is Gaussian if $\L_{lm}(x)\equiv0$; for 
$\varphi(x)$ to be non-Gaussian requires $\L_{lm}(x)\not\equiv0$. 

Fields with truncated correlations of the form (7) can have no conventional 
particle scattering. For example, with $\varphi(h)\equiv\tint 
\varphi(x)\,h(x)\,d^n\!x$, it follows from (7) that 
$\<0|\v(h_1)\v(h_2)\v(h_2)\v(h_1)|0\>^T\ge0$, and by an associated 
Schwarz inequality \cite{buc} that
 \bn &&\hskip-1cm0\le|\<0|\v(f_1)\v(f_2)\v(g_2)\v(g_1)|0\>^T|^2 \n \\
 &&\le \<0|\v(f_1)\v(f_2)\v(f_2)\v(f_1)|0\>^T\,\<0|\v(g_1)\v(g_2)
\v(g_2)\v(g_1)|0\>^T\;. \en
Passing to asymptotic fields, $\v(f)\ra\v_{\rm out}(f)$, and 
$\v(g)\ra\v_{\rm in}(g)$, leads to 
\bn  &&\hskip-.6cm0\le|\<0|\v_{\rm out}(f_1)\v_{\rm out}(f_2)
\v_{\rm in}(g_2)\v_{\rm in}(g_1)|0\>^T|^2 \n \\
 &&\hskip-.6cm\le \<0|\v_{\rm out}(f_1)\v_{\rm out}(f_2)\v_{\rm out}
(f_2)\v_{\rm out}(f_1)|0\>^T\,\<0|\v_{\rm in}(g_1)\v_{\rm in}
(g_2)\v_{\rm in}(g_2)\v_{\rm in}(g_1)|0\>^T\n\\  &&\hskip1cm =0\;. \en
This behavior is consistent with the assumption that 
$\L_{lm}(x)\ra\L^{{\rm out},{\rm in}}_{lm}(x)\equiv0$ and 
$C_l(x)\ra C_l^{\rm out}(x)=C^{\rm in}_l(x)$.

On the surface, it would appear that a quantum theory with no conventional 
scattering 
would be inconsistent with the original classical theory which is known 
to exhibit nontrivial classical scattering \cite{rei}. However, an 
interchange of limits is involved here, which, on closer inspection, 
shows that no inconsistency arises.

If we assume the suggested form for the nonclassical, auxiliary 
potential $P$, there exists, in effect, an additional term in the 
operator energy density proportional to $\hbar^2/\v(x)^2$, or in the 
corresponding Heisenberg equation of motion an additional term 
proportional to $\hbar^2/\v(x)^3$. Such terms {\it preclude} the 
field operator from reaching arbitrarily small values, and hence 
nontrivial asymptotic fields (with $\hbar>0$) fail to exist in the 
usual sense.

We summarize the situation as follows: On the one hand, we observe that 
taking the classical limit ($\hbar\ra0$) after going to the asymptotic 
quantum fields leaves us with no classical scattering. On the other 
hand, if one takes the classical limit ($\hbar\ra0$) of the quantum 
equations of motion at finite time values and solves the resultant 
classical theory, then one should find the nontrivial scattering of the 
classical theory alluded to before.

>From another perspective, the behavior just outlined can also lead to 
intermediate situations. Let us imagine a quantum system endowed with 
a great deal of energy and composed of a huge number of quanta. By the 
correspondence principle, the associated scattering behavior can be 
approximately treated classically and is therefore nonvanishing provided 
one sticks to possibly large, but {\it finite}, preparation and detection 
times in the past and future, respectively. In other words, such a field 
may not exhibit conventional quantum particle scattering in the strict 
sense, but under suitable conditions, an effective scattering theory 
may well exist that could help in recovering the nontrivial scattering 
of the original classical theory. Moreover, should there be any stable 
large-field configurations, such as solitons, then scattering between 
such entities may well exist.

With $A(a)\ra\infty$ sufficiently fast, the proposed form of $P$ given 
in (5) can avoid the fate---irrelevancy---of usual higher-order interactions 
in a renormalization-group treatment. Any analysis of these models should 
perhaps begin with the case $g_o\equiv 0$ for which, thanks to $P$, 
nontriviality is still expected.

\end{document}